\documentclass[5p, times]{elsarticle}
\usepackage[utf8]{inputenc}

\usepackage{amssymb}
\usepackage{amsthm}
\usepackage{amsmath}
\usepackage{hyperref}
\usepackage{lineno}
\usepackage{tikz}
\usetikzlibrary{positioning, arrows.meta}
\usepackage{enumitem}
\usepackage{fancyvrb}
\usepackage{listings}

\definecolor{codegreen}{rgb}{0,0.6,0}
\definecolor{codegray}{rgb}{0.5,0.5,0.5}
\definecolor{codepurple}{rgb}{0.58,0,0.82}
\definecolor{backcolour}{rgb}{1., 1., 1.}

\lstdefinestyle{mystyle}{
    backgroundcolor=\color{backcolour},   
    commentstyle=\color{codegreen},
    keywordstyle=\color{magenta},
    numberstyle=\tiny\color{codegray},
    stringstyle=\color{codepurple},
    basicstyle=\ttfamily\footnotesize,
    breakatwhitespace=false,         
    breaklines=true,                 
    captionpos=b,                    
    keepspaces=true,                 
    numbers=left,                    
    numbersep=5pt,                  
    showspaces=false,                
    showstringspaces=false,
    showtabs=false,                  
    tabsize=2,
    abovecaptionskip=10pt
}

\newcommand{\package}[1]{\textit{#1}}
\newcommand{\ixpeobssim}{\package{ixpeobssim}}
\newcommand{\app}[1]{\textit{#1}}
\newcommand{\col}[1]{\texttt{#1}}
\newcommand{\secref}[1]{\S\ref{sec:#1}}
\newcommand{\aeff}{\ensuremath{A_{\text{eff}}}}
\newcommand{\modf}{\ensuremath{\mu}}
\newcommand{\Eqref}[1]{Eq.~\eqref{#1}}

\begin{document}

\begin{frontmatter}

\title{\ixpeobssim: a Simulation and Analysis Framework for the Imaging X-ray Polarimetry Explorer}

\author[1,2]{Luca Baldini}
\author[3,4,5]{Niccol\`o Bucciantini}
\author[6]{Niccol\`o Di Lalla}
\author[7]{Steven Ehlert}
\author[2]{Alberto Manfreda}
\author[8,9,10]{Michela Negro}
\author[6]{Nicola Omodei}
\author[2]{Melissa Pesce-Rollins}
\author[2]{Carmelo Sgr\`o}
\author[1,2]{Stefano Silvestri}
\address[1]{Università di Pisa, Dipartimento di Fisica Enrico Fermi,
    Largo B. Pontecorvo 3, I-56127 Pisa, Italy}
\address[2]{Istituto Nazionale di Fisica Nucleare, Sezione di Pisa, 
    Largo B. Pontecorvo 3, I-56127 Pisa, Italy}
\address[3]{Istituto Nazionale di Astrofisica, Osservatorio Astrofisico di Arcetri,
    Largo E. Fermi 5, I-50125, Firenze, Italy}
\address[4]{Dipartimento di Fisica \& Astronomia, Univesit\`a degli Studi di Firenze,
    Via G. Sansone 1, 50019, Sesto F.no, Italy}    
\address[5]{Istituto Nazionale di Fisica Nucleare, Sezione di Firenze, Via G. Sansone 1, 50019 Sesto F.no, Italy}
\address[6]{W.W. Hansen Experimental Physics Laboratory, Kavli Institute for Particle Astrophysics and Cosmology, Department of Physics and SLAC National Accelerator Laboratory, Stanford University, Stanford, CA 94305, USA}
\address[7]{NASA Marshall Space Flight Center, Huntsville, AL 35812, USA}
\address[8]{University of Maryland, Baltimore County, Baltimore, MD 21250, USA}
\address[9]{NASA Goddard Space Flight Center, Greenbelt, MD 20771, USA}
\address[10]{Center for Research and Exploration in Space Science and Technology, NASA/GSFC, Greenbelt, MD 20771, USA}

\journal{SoftwareX}
\date{Compiled on \today}


\begin{abstract}

\ixpeobssim\ is a simulation and analysis framework specifically developed for the Imaging X-ray
Polarimetry Explorer (IXPE). Given a source model and the response functions of the telescopes,
it is designed to produce realistic simulated observations, in the form of event lists in FITS 
format, containing a strict superset of the information included in the publicly released
IXPE data products. The core simulation capabilities are complemented by a full suite of 
post-processing applications which support the spatial, spectral, and temporal models 
needed for analysis of typical polarized X-ray sources, allowing for the implementation of complex,
polarization-aware analysis pipelines, and facilitating the interoperation with the standard 
visualization and analysis tools traditionally in use by the X-ray community.
Although much of the framework is specific to IXPE, the modular nature of the underlying 
implementation makes it potentially straightforward to adapt it to different missions with 
polarization capabilities.

\end{abstract}


\end{frontmatter}

\section{Introduction}
\label{sec:introduction}

Launched on December 9, 2021, the Imaging X-ray Polarimetry Explorer (IXPE) is a NASA 
Small Explorer Mission developed in collaboration with the Italian Space 
Agency~\cite{weisskopf2021imaging, Soffitta_2021, BALDINI2021102628}, and the first ever
to provide position-resolved polarimetric capabilities in the 2--8~keV energy band.

IXPE recovers the linear polarization of the source on a statistical basis by measuring the
azimuthal distribution of the photo-electrons generated by X-rays absorbed in the detector,
and complements the polarization sensitivity with timing, spectral and imaging capabilities~\cite{weisskopf2021imaging}. From the standpoint of high-level science analysis,
each event is characterized by five independent quantities: the arrival time, the energy, two sky-coordinates and the azimuthal angle $\phi_i$ of the photo-electron in the tangent-plane 
projection. This information is complemented by an additional quantity representing the
estimated quality of the direction reconstruction (or \emph{weight}) $w_i$, that can be 
exploited in an ensemble analysis to enhance the polarization sensitivity~\cite{dimarco2022weighted}.

Rather than using $\phi_i$ directly, we encode the polarization information in two event-by-event
Stokes parameters
\begin{align}
    q_i = 2 \cos 2\phi_i \quad \text{and} \quad u_i = 2 \sin 2\phi_i,
\end{align}
following the formalism in~\cite{KISLAT201545}%
\footnote{Note the extra factor of $2$ with respect to the original paper.}.
This is primarily due to the approach used for calibrating the detector response
to un-polarized radiation~\cite{Rankin_2022}: the event-by-event Stokes parameters after the 
subtraction of the spurious modulation are no longer properly normalized, and (although the effect
is irrelevant in any large ensemble of events) cannot be readily interpreted in terms of azimuthal 
angles. In addition, a formulation in Stokes parameter space is less prone to the pitfalls associated
to the fact that the polarization angle and degree, unlike the Q and U Stokes parameters,
are not statistically independent.

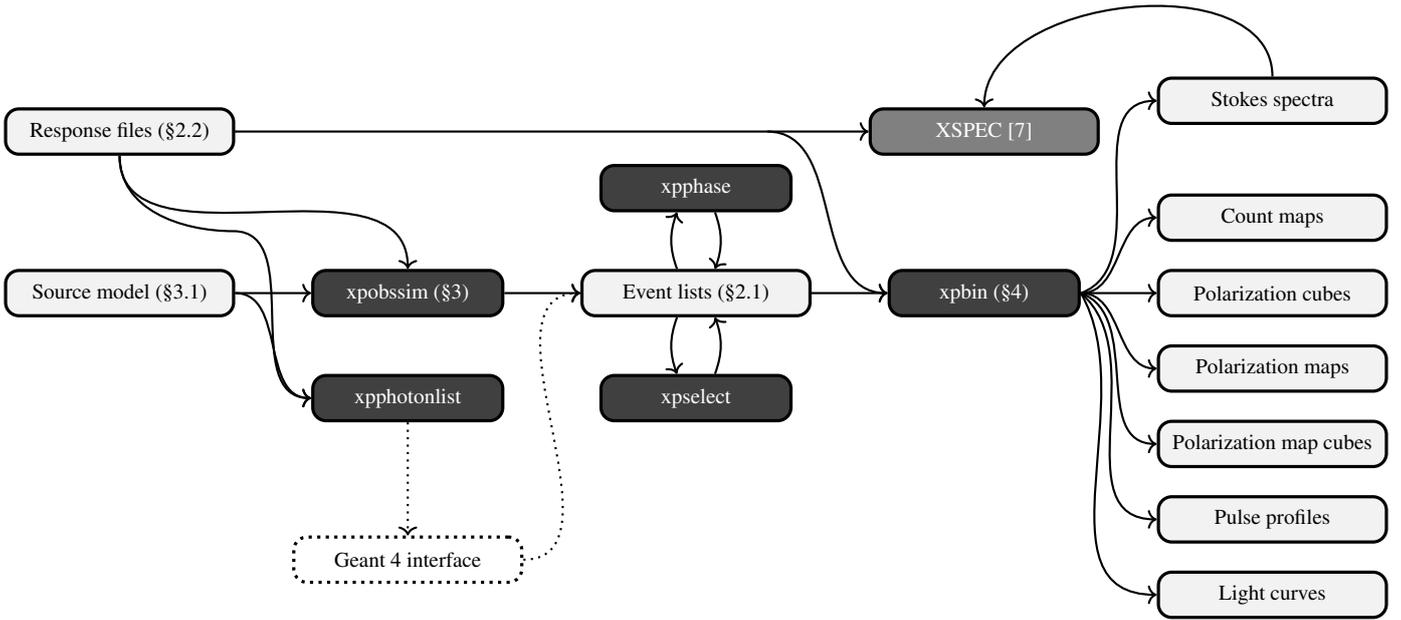
\begin{figure*}[htb!]
    \centering
    \begin{tikzpicture}[node distance=1.5cm and 1.cm,
    prod/.style={rectangle, draw=black, fill=black!5, very thick, minimum width=3.cm,
        minimum height=0.6cm, rounded corners=5pt},
    app/.style={rectangle, draw=black, fill=black!75, text=white, very thick, minimum width=2.5cm, 
        minimum height=0.6cm, rounded corners=5pt},
    extapp/.style={rectangle, draw=black, fill=black!50, text=white, very thick, minimum width=3.cm, 
        minimum height=0.6cm, rounded corners=5pt},
    privapp/.style={rectangle, draw=black, very thick, minimum width=3.cm, dotted,
        minimum height=0.6cm, rounded corners=5pt},
    conn/.style={-{Computer Modern Rightarrow}, thick}
    ]
    \footnotesize
    
    \node[prod](respfiles){Response files (\secref{caldb})};
    \node[prod](srcmodel)[below=of respfiles]{Source model (\secref{srcmodel})};
    \node[app](xpobssim)[right=of srcmodel]{xpobssim (\secref{simulation})};
    \node[app](xpphotonlist)[below=of xpobssim, yshift=0.75cm]{xpphotonlist};
    \node[privapp](g4)[below=of xpphotonlist]{Geant~4 interface};
    \node[prod](evtlist)[right=of xpobssim]{Event lists (\secref{datafmt})};
    \node[app](xpbin)[right=of evtlist]{xpbin (\secref{analysis})};
    \node[prod](stokesspec)[right=of xpbin, yshift=2.55cm]{Stokes spectra};
    \node[prod](cmap)[right=of xpbin, yshift=1cm]{Count maps};
    \node[prod](pcube)[right=of xpbin, yshift=0cm]{Polarization cubes};
    \node[prod](pmap)[right=of xpbin, yshift=-1cm]{Polarization maps};
    \node[prod](pmapcube)[right=of xpbin, yshift=-2cm]{Polarization map cubes};
    \node[prod](pp)[right=of xpbin, yshift=-3cm]{Pulse profiles};
    \node[prod](lc)[right=of xpbin, yshift=-4cm]{Light curves};
    \node[app](xpselect)[below=of evtlist, yshift=0.75cm]{xpselect};
    \node[app](xpphase)[above=of evtlist, yshift=-0.75cm]{xpphase};
    \node[extapp](XSPEC)[above=of xpbin]{XSPEC~\cite{1996ASPC..101...17A}};
    \draw[conn] (respfiles.south) to[out=-90, in=90] (xpobssim.north);
    \draw[conn] (respfiles.south) to[out=0, in=90] (respfiles.south) + (1.5cm, -1cm)
        to[out=180, in=-90] (respfiles.south) + (1.5cm, -1cm) to[out=0, in=180] (xpphotonlist.west);
    \draw[conn] (srcmodel.east) to (xpobssim.west);
    \draw[conn] (srcmodel.east) to[out=0, in=180] (xpphotonlist.west);
    \draw[conn] (xpobssim.east) to (evtlist.west);
    \draw[conn] (evtlist.east) to (xpbin.west);
    \draw[conn] (xpbin.east) to[out=20, in=180] (stokesspec.west);
    \draw[conn] (xpbin.east) to[out=0, in=180] (cmap.west);
    \draw[conn] (xpbin.east) to[out=0, in=180] (pcube.west);
    \draw[conn] (xpbin.east) to[out=0, in=180] (pmap.west);
    \draw[conn] (xpbin.east) to[out=-20, in=180] (pmapcube.west);
    \draw[conn] (xpbin.east) to[out=-40, in=180] (pp.west);
    \draw[conn] (xpbin.east) to[out=-60, in=180] (lc.west);
    \draw[conn] (stokesspec.north) to[out=90, in=90] (XSPEC.north);
    \draw[conn] (respfiles.east) to (XSPEC.west);
    \draw[conn, transform canvas={xshift=0.25cm}, bend right=20] (xpselect.north) to (evtlist.south);
    \draw[conn, transform canvas={xshift=-0.25cm}, bend right=20] (evtlist.south) to (xpselect.north);
    \draw[conn, transform canvas={xshift=0.25cm}, bend left=20] (xpphase.south) to (evtlist.north);
    \draw[conn, transform canvas={xshift=-0.25cm}, bend left=20] (evtlist.north) to (xpphase.south);
    \draw[conn] (respfiles.east) to (respfiles.east) + (7cm, 0cm)
        to[out=0, in=180] (xpbin.west);
    \draw[conn, dotted] (xpphotonlist.south) to (g4.north);
    \draw[conn, dotted] (g4.east) to[out=0, in=180] (evtlist.west);
    \end{tikzpicture}
    \caption{Simplified \ixpeobssim\ architectural overview. The dark boxes identify specific \ixpeobssim\ 
    applications, whereas light boxes represent different types of data products. (XSPEC, as an 
    external program, is rendered in a different style.) The numbers in parentheses provide
    the reference to the proper section of the paper where each item is discussed.
    The interface to the Geant~4~\cite{AGOSTINELLI2003250} detector simulation, which 
    will be briefly discussed in section~\ref{sec:weights}, is rendered with a dotted line style
    beacause it is not part of the publlic release.}
    \label{fig:ixpeobssim_overview}
\end{figure*}

In this paper we describe \ixpeobssim, a framework designed to simulate IXPE observations
of celestial sources, producing event lists in a FITS format that is intended to be a strict
superset of that of the standard data products used for science analysis.
To our best knowledge, the polarization-specific functionalities provided by \ixpeobssim\ 
are not readily available in any of the existing X-ray simulation toolkits tailored to
imaging, spectroscopy and/or timing---be they mission-specific, such as \package{MARX}~\cite{marx},
\package{NuSim}~\cite{nusim} or the now discontinued \package{SciSim}~\cite{SciSim},
or general-purpose, such as \package{simx}~\cite{simx} and \package{SIXTE}~\cite{sixte}.

In addition to the simulation facilities, \ixpeobssim\ provides a set of applications to filter, 
reduce, analyze and visualize both simulated and real data, which we anticipate will be a useful
resource for the community engaged in the analysis of IXPE data.
To this end, we shall briefly discuss, where appropriate, the interplay and the overlap
with the software tools and data products released by the HEASARC in support of the IXPE mission,
and we shall emphasize the specific functionalities that are peculiar to \ixpeobssim.

\section{Architectural Overview}
\label{sec:overview}

The \ixpeobssim\ framework is based on the Python programming language and makes extensive
use of the associated scientific ecosystem, most notably \package{numpy}~\cite{harris2020array},
\package{SciPy}~\cite{2020SciPy-NMeth} and \package{matplotlib}~\cite{Hunter:2007},
as well as the de-facto standard package for numerical analysis in astronomy:
\package{Astropy}~\cite{astropy:2013, astropy:2018}.
Leveraging the Python intuitive syntax, extensibility and introspection capabilities, 
\ixpeobssim\ is streamlined for speed and modularity, with the ultimate goal of making it 
easy for the user to create complex simulations and analysis workflows.
The lack of an X-ray simulation framework written in a modern high-level, interpreted  
language was in fact one of the main motivations for us to develop one from scratch, 
rather than building upon the aforementioned existing
toolkits~\cite{marx, nusim, SciSim, simx, sixte}.

Figure~\ref{fig:ixpeobssim_overview} shows a simplified architectural overview of the 
framework. The main application, \app{xpobssim}, takes a complete source model (including 
the temporal, morphological, spectral and polarimetric characteristics) and a coherent 
set of parameterized response files---most notably the effective area and associated
vignetting function, the response matrix, the modulation factor as a function of the energy,
as well as a model of the point-spread function (PSF)---to produce an event list that is 
germane to an actual file from a celestial observation. The construction of the source models, 
the response files and the  format of the event lists will be covered in more detail in 
sections~\ref{sec:srcmodel},  \ref{sec:caldb} and \ref{sec:datafmt}, respectively.

\ixpeobssim\ provides tools for modifying and post-processing event lists, among which 
\app{xpselect} allows to apply arbitrary selections, e.g., on time, energy 
and  position in the sky. Filtered event lists are intended to be functionally identical to their
parents to be able to inter-operate with all the analysis tools in exactly the same fashion.

\app{xpbin} provides the capability of reducing event lists, generating binned data 
products in a number of different fashions, including \emph{Stokes spectra} 
(that is, OGIP-compliant, type 1 PHA files~\cite{ogip_92_007} for $I$, $Q$ and $U$, 
in detector space, suitable to perform spectro-polarimetric fits as described  
in~\cite{Strohmayer_2017}) as well as several different data structures encapsulating
the results of a model-independent ensemble analysis a la~\cite{KISLAT201545}
(this will be further discussed in section~\ref{sec:pol_analysis}).

We note that the HEASOFT \package{xselect} FTOOL provides support for part of the 
same functionality since version 6.30---more specifically for filtering IXPE event lists and 
creating binned Stokes spectra. At this time, we consider the benefits deriving from the full 
integration with the rest of package to outweigh the cost of the limited duplication.
As \ixpeobssim\ evolves, we shall strive to adapt its public interfaces to make them as similar
as possible to the HEASARC FTOOL equivalent;  the \ixpeobssim\ documentation includes a 
comprehensive and  up-to-date description of the interplay with HEASOFT.

\subsection{Output data format}
\label{sec:datafmt}

The standard IXPE high-level data products distributed by the HEASARC include:
\begin{enumerate}[leftmargin=10pt]
    \item level-1 FITS files containing individual photoelectron track images and associated
    reconstructed quantities in detector coordinates;
    \item level-2 filtered event lists providing all the physical information in sky coordinates
    that is relevant for science analysis~\cite{soc}.
\end{enumerate}
Event lists generated by \ixpeobssim\ include the two extensions of the standard level-2 file:
\col{EVENTS}, containing the event data, and \col{GTI}, listing the good time intervals
for the observation. The \col{EVENTS} extension for simulated event contains a few columns for
diagnostic purposes (e.g., event positions in detector coordinates) that for real observations
are only included in level-1 files---and none of which is used by the high-level analysis tools.

The ground truth, including the true photon energy, the true direction in the sky and a
unique identifier of the particular source (or source component) in the field of view that
originated the event, is captured in a dedicated \col{MONTE\_CARLO} extension.
The latter is useful for debugging purposes, and because it readily provides a way
to evaluate the effect of the detector response on the relevant high-level observables.
These include, e.g., the polarization dilution effect due to the finite angular resolution
for extended sources where the polarization angle varies over spatial scales comparable
to the PSF, and the effect of the energy dispersion on high-level polarization analysis.
A few additional binary extensions can be optionally generated, and will be briefly described 
in the following sections.

\subsection{Response functions: the pseudo calibration database}
\label{sec:caldb}

The instrument response functions (IRFs) are a fundamental part of \ixpeobssim, and they are
used, in identical form, for both the simulation and the science analysis. Notably,
this allows for a comprehensive verification of the full analysis workflow under 
controlled conditions. All the response files are OGIP-compliant and are intended to be
inter-operable with the analysis tools provided by HEASARC.

More specifically, we identify six different types of response functions: on-axis 
effective area and response matrix (stored in standard .arf and .rmf FITS files),
vignetting, point-spread function, modulation factor and modulation response function. 
The modulation factor represents the response of the detector to 100\% polarized radiation,
and serves the purpose of converting the azimuthal modulation measured by the detector
into the actual source polarization. The modulation response function is the product
of the modulation factor and the on-axis effective area, and is used as the proper
ancillary response files for $Q$ and $U$ spectra in polarimetric fits, as explained 
in section~\ref{sec:pol_analysis}.

\ixpeobssim\ provides facilities for generating, reading, displaying and using IRFs.
Due to the specific needs of the simulation facilities, \ixpeobssim\ is distributed with 
its own, self-contained calibration database, that we shall refer to as the
\emph{pseudo-CALDB} and is structurally equivalent to the real database distributed 
through HEASARC. 
All the response files that are relevant for science analysis (i.e., effective 
area, response matrix, and modulation response files) are properly synchronized between 
the two databases, and a dedicated table mapping the correspondence between the 
pseudo-CALDB and the real CALDB is maintained as part of the \ixpeobssim\ documentation.

\section{Simulating Observations}
\label{sec:simulation}

\subsection{Source model definition}
\label{sec:srcmodel}

The basic characteristics of each model \emph{component} are specified as ordinary Python
functions: the photon spectrum can be an arbitrary function of energy and time (or phase,
for periodic sources), while the polarization degree and angle can be an arbitrary function 
of energy, time (or phase), and sky-direction:
\begin{align}\label{eq:srcmod}
    \begin{cases}
    \mathcal S (E, t)  & [\text{cm}^{-2}~\text{s}^{-1}~\text{keV}^{-1}]\\
    P_D (E, t, x, y)   & \\
    P_A (E, t, x, y)   & [\text{rad}].
    \end{cases}
\end{align}
This approach allows for a large degree of flexibility, as the function bodies can contain,
e.g., complex analytic functions or interpolators built from numerical tables, provided that
the signature is correct.
We emphasize that, since the input model does not have associated errors by its nature, 
our treatment is strictly equivalent to a formulation in Stokes parameter space, and none
of the nuisances connected with the fact that \emph{measured} polarization degree and angle
are not independent applies to the simulation process.

\ixpeobssim\ supports a wide range of models for source morphology via a hierarchy
of classes describing point sources, simple geometrical shapes like disks or annuli, and 
arbitrary extended sources based on intensity maps in FITS format.
Extended sources defined in this way come with the limitation that the input image 
only controls the normalization of the spectrum, while spectral shapes explicitly depending 
on the position cannot be specified in~\Eqref{eq:srcmod}---although one can always tessellate
the source with an arbitrary number of independent patches.
Alternatively, an interface to Chanda ACIS S/I event lists allows defining model components 
that can take advantage of the superior angular and energy resolution of this observatory
with respect to IXPE and provides an alternative simulation strategy fully preserving the
correlations between the position in the sky and the spectral shape, overcoming the seemingly
restrictive constraints on the signature of the photon spectrum.

The concept of \emph{model component} is used throughout \ixpeobssim\ to indicate a number
of different objects, ranging from a simple celestial source to different physical components
of the same source (e.g., the thermal and non-thermal emission), or different physical sources
in the same field (e.g., a PWN and its pulsar). The basic simulation unit, that we refer to as a
\emph{region of interest} (ROI), is a collection of an arbitrary number of model components 
within the IXPE field of view, encapsulated in a Python configuration file that can
be fed into \app{xpobssim}. A minimal working example is listed in~\ref{app:code}.

\subsection{Simulation workflow}

\begin{figure}[!htb]
    \centering
    \includegraphics[width=\linewidth]{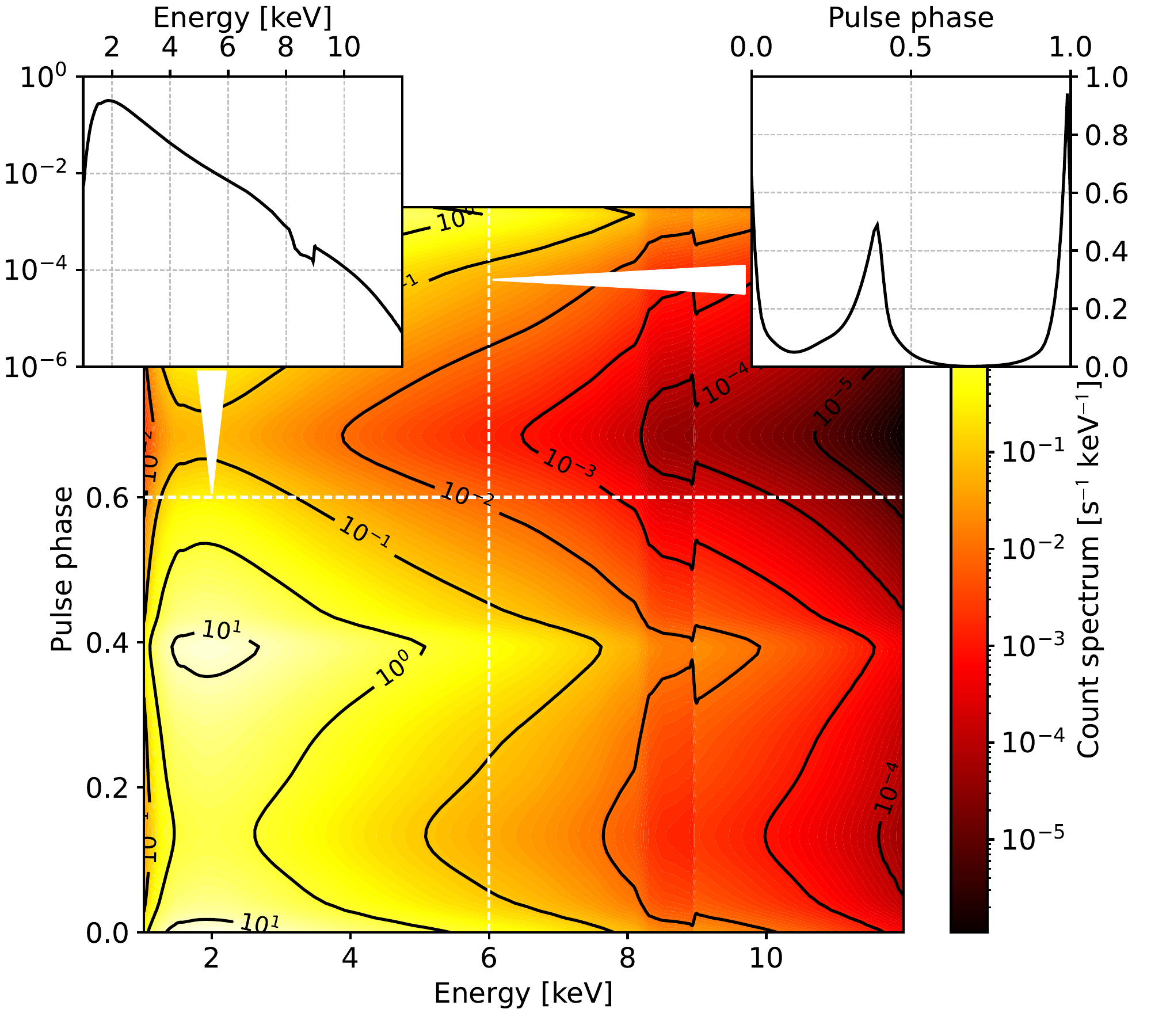}
    \caption{Representation of a simple model for the count spectrum of the Crab pulsar
    (note that, because the source is periodic, the $y$-axis represents the pulse phase).
    Any horizontal slice represents the photon spectrum, convolved with the on-axis
    effective area of the telescope, at a given phase, while any vertical slice represents
    the pulse profile at a given energy.}
    \label{fig:crab_count_spectrum}
\end{figure}

Considering an on-axis point source for simplicity, the basic flow of the simulation for a 
single model component starts with the calculation of the count spectrum, as a function of
energy and time (or phase), given the photon spectrum $\mathcal{S}(E, t)$ and the on-axis 
effective area $A_{\rm eff}(E)$:
\begin{align}
    \mathcal{C}(E, t) = \mathcal{S}(E, t) \times A_{\rm eff}(E) \quad
    [\text{s}^{-1}~\text{keV}^{-1}].
\end{align}
A simple model for the phase-resolved count spectrum of the Crab pulsar is shown in
Figure~\ref{fig:crab_count_spectrum} for illustrative purposes.

The light curve (or the pulse profile) in counts space is readily obtained by integrating
over the energy
\begin{align}
    \mathcal{L}(t) = \int_{E_{\rm min}}^{E_{\rm max}} \mathcal{C}'(E, t) \, \mathrm{d}E \quad [\text{s}^{-1}], 
\end{align}
which in turn allows to calculate the total number of expected events $N_{\rm exp}$ 
by integrating over the observation time:
\begin{align}
    N_{\rm exp} = \int_{t_{\rm min}}^{t_{\rm max}} \mathcal{L}(t) \, \mathrm{d}t.
\end{align}

We extract the number of observed events $N_{\rm obs}$ according to a Poisson distribution 
with mean $N_{\rm exp}$ and treat the light curve as a one-dimensional probability 
density function (pdf) to extract the initial vector of event times $t_i$ (or phases $p_i$, 
for periodic sources). For each event we use the count spectrum $\mathcal{C}(E, t_i)$, 
calculated at the proper time or phase, as a one-dimensional pdf from which we extract 
the true energy $E_i$. 
The true sky-direction $(x_i, y_i)$ is extracted independently---more specifically, 
each class in the model hierarchy provides a specific implementation of a dedicated base
method, encapsulating the proper sampling of the underlying model~\eqref{eq:srcmod}, at
the given $E_i$ and $t_i$.

The basic steps outlined above complete the generation of the ground truth for the
simulation, which is then convolved with the instrument response. In this respect, the 
\ixpeobssim\ implementations of data structures encapsulating the point-spread function and
the energy dispersion readily provide facilities for efficient random sampling of the
corresponding probability density functions, which are used to extract the \emph{measured}
energy and direction in the sky.

At this point of the simulation workflow we also have all the necessary ingredients
to extract the photo-electron emission angle, according to the proper azimuthal distribution
\begin{align}
    p_\phi(\phi; m, \phi_0) = \frac{1}{2\pi} \left\{ 1 + m \cos(2(\phi - \phi_0)) \right\},
\end{align}
where the visibility of the modulation is given by the product of the polarization 
degree $P_D$ of the underlying model~\eqref{eq:srcmod} and the modulation factor, calculated
at the event energy $E_i$
\begin{align}
    m_i = P_D(E_i, t_i, x_i, y_i) \times \mu(E_i).
\end{align}
The phase $\phi_0$, corresponding to the position angle $P_A$ in~\Eqref{eq:srcmod},
can then be added, provided that the result is folded back into the desired interval $[-\pi, \pi]$.

The process is repeated independently for all the source components in the ROI, and the partial
event lists are then merged and sorted in time. All the additional instrumental
corrections are applied in dedicated filtering stages before the final, consolidated
event list is written to file. More specifically:
\begin{itemize}[leftmargin=10pt]
    \item the events are initially generated using the on-axis effective area, and the
    correct vignetting function is obtained by randomly discarding a fraction of the
    events with the proper dependence on the off-axis angle;
    \item events with a projected position on the focal plane lying outside the fiducial 
    area of the readout chip are discarded;
    \item events occurring during the readout of an earlier event are discarded, as well.
\end{itemize}

\subsection{Low-level implementation}

The vast majority of the pdfs involved in the simulation of celestial sources can only be sampled
via numerical methods---even a simple power-law photon spectrum, when convolved with the telescope 
effective area, cannot be sampled by analytical means. We largely rely on interpolating splines
as an effective means to sample arbitrary random variables.

For one-dimensional pdfs, we calculate the cumulative function on a suitable, regular grid
and use the values to build an interpolated spline representing the percent point function
(ppf), that can in turn be used to sample the underlying random variable $r$ via
inverse transform~\cite{pdg}:
\begin{align}
    r = \text{ppf}(\xi) \quad \text{where} \quad \xi \sim \text{Uniform}(0,~1).
\end{align}
Provided that the order of the spline and grid spacing are selected properly for the 
situation at hand, this allows for a formulation of the problem that can be naturally
vectorized in an efficient fashion through the \package{numpy} facilities. 
Implementation details aside, this is the basic strategy that we use to sample the event
times from the light curve (or the phase values from the pulse profile for periodic sources).

\begin{figure}[!htb]
    \centering
    \includegraphics[width=\linewidth]{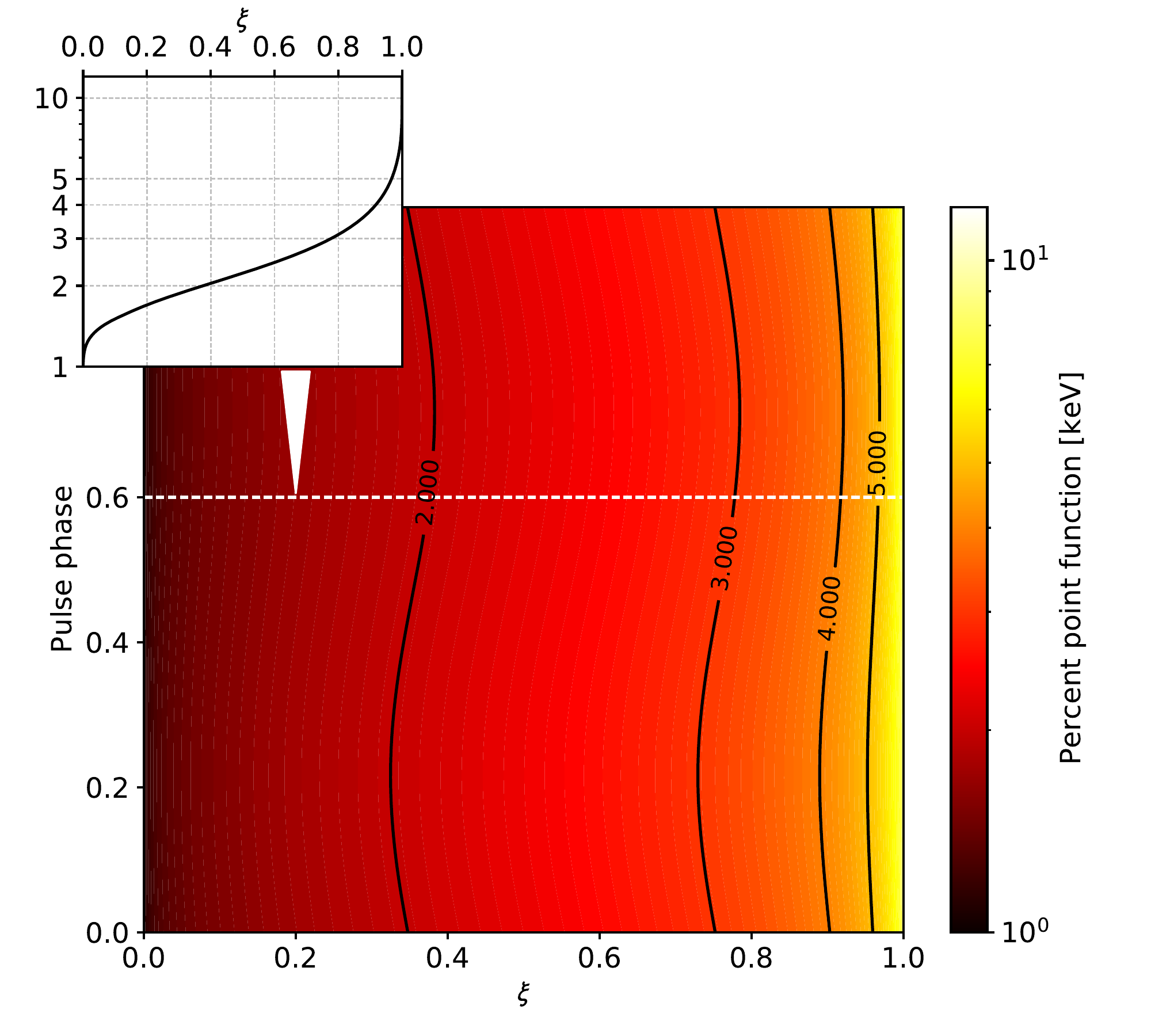}
    \caption{Bi-dimensional ppf corresponding to the count Crab spectrum in
    Figure~\ref{fig:crab_count_spectrum}. Each horizontal slice represents the actual ppf at 
    a given pulse phase, and can be used to sample the underlying event energy in a fully
    vectorized fashion.}
    \label{fig:crab_ppf}
\end{figure}

The problem of extracting the event energies given a count spectrum is intrinsically 
more complex, due to the fact that the spectral shape, in general, depends on the 
event time or phase. As re-computing the one-dimensional pdf for each event would be overly
computationally-intensive, we resort to pre-calculating the ppf values on a regular grid
of time (or phase) values and use them to build an interpolated bi-variate spline, that 
we refer to as the \emph{horizontal} ppf (hppf), as illustrated
in~Figure~\ref{fig:crab_ppf}. The latter can then be used to sample the 
underlying 2-dimensional distribution in a vectorized fashion:
\begin{align}
    E = \text{hppf}(\xi, t)  \quad \text{where} \quad \xi \sim \text{Uniform}(0,~1).
\end{align}
This is the basic approach that, properly re-cast in azimuthal angle-modulation space, 
we also use to sample the emission direction of the photo-electrons.
It is worth noting that being able to specify a source model in terms of a 
position-dependent spectrum $\mathcal S(E, t, x, y)$ would require generalizing this
method from two to four dimensions, which is the main reason for the corresponding
limitation in~\Eqref{eq:srcmod}.

\subsection{Event weights}
\label{sec:weights}

Since the the \ixpeobssim\ simulation facilities are entirely based on a parametrization 
of the detector response, as opposed to an actual microscopic simulation of particle
interactions, one of the intrinsic limitations is the inability to generate track images.
This, in turn, makes it non trivial to properly simulate the event-by-event weights and 
provide the complete set of information that one would find in real data.

To this end, the IXPE Collaboration has been largely relying on the \ixpeobssim\ interface
to the microscopic Monte Carlo simulation of the detectors, based on the Geant~4
toolkit~\cite{AGOSTINELLI2003250}, developed to support the design and implementation
of the mission and inform the generation of the response files. This provides the ultimate
fidelity (at the expense of a much longer simulation time) and has been extensively used
to test the use of weights in the tools that \ixpeobssim\ provides to analyze real
observations. However, due to its inherent complexity, as well as the dependence of
various large external libraries, packaging the detector simulation into a form that could
be publicly distributed and supported was deemed to be too resource consuming, given the
relatively niche use case.

At the time of writing, the only \emph{pseudo-weighted} workflow that \ixpeobssim\
supports is to generate event lists using the weighted response functions---with the weights
being automatically set to 1 for all the events. From a sensitivity standpoint, this simplistic
approach provides the exact equivalent of a fully-fledged, weighted analysis, and guarantees
that simulated data are transparently inter-operable with all the high-level analysis tools,
but is not sufficient for an end-to-end test of the entire simulation and analysis chain.
This is currently one of the main limitations of the package.

Future \ixpeobssim\ versions might conceivably offer realistic event weights through 
a hybrid approach, where we use the full detector simulation to generate a static,
three-dimensional lookup table parametrized in the pulse invariant--true energy phase space
to be used at simulation time to sample the weight values.

\subsection{Advanced source models}

\ixpeobssim\ provides a comprehensive set of high-level interfaces to facilitate the 
coding of complex source models. Dedicated Python facilities allow to compose spectral
models with time- or phase-dependent parameters, e.g., a power law where the normalization 
and/or the spectral index are functions of time or pulse phase.
Arbitrary \package{XSPEC} spectral models can be automatically wrapped into the proper 
function signature and fed natively into \app{xpobssim}, building on top of the
\package{PyXSPEC} Python interface.
In addition, a fully-fledged interface to OGIP-compliant libraries of tabular fitting
models, with interpolation capabilities, is available---and a real-life example, interfacing
to the magnetar models described in~\cite{10.1093/mnras/staa204}, is provided.
Finally, complex polarization patterns for extended sources can be specified via a
dedicated data structure encapsulating a collection of sky-maps of Stokes parameters in
different energy layers, leveraging the \package{SciPy} capability of interpolating
on a regular grid in an arbitrary number of dimensions.

A comprehensive review of the \ixpeobssim\ modeling facilities is beyond the scope of this
paper, and the subject is largely covered in the documentation, to which the reader is referred.

\subsection{Good time intervals}

IXPE operates in an approximately circular low Earth orbit at 601.1~km altitude and 
$0.23^\circ$ inclination. \ixpeobssim\ captures the main features of the motion of the 
spacecraft around the Earth by means of a representative two-line element (TLE) set 
based on observations of the spacecraft taken soon after launch

There are three main ways in which the spacecraft orbit needs to be accounted for in
simulated data: the Earth can obstruct the line of sight between the spacecraft and
observation target; no observations can take place while the spacecraft is above the South
Atlantic Anomaly (SAA); and certain celestial positions are only available for observations
at certain times of the year due to Sun angle constraints. These values can be of crucial
importance when trying to coordinate observations with other telescopes or accounting for
the expected gaps in coverage for a time or pulse phase-dependent polarimetry signal. 
We use the \package{Skyfield}~\cite{2019ascl.soft07024R} package to geo-locate 
the spacecraft as a function of time -- which, in turn, enables all three of these effects
to be incorporated into the production of Good Time Intervals (GTI) for simulated observations.
Although they are not identical to those expected in a real observation, they are
statistically representative of the latter.

\subsection{Dithering and pointing history}

\begin{figure}[!htb]
    \centering
    \includegraphics[width=\linewidth]{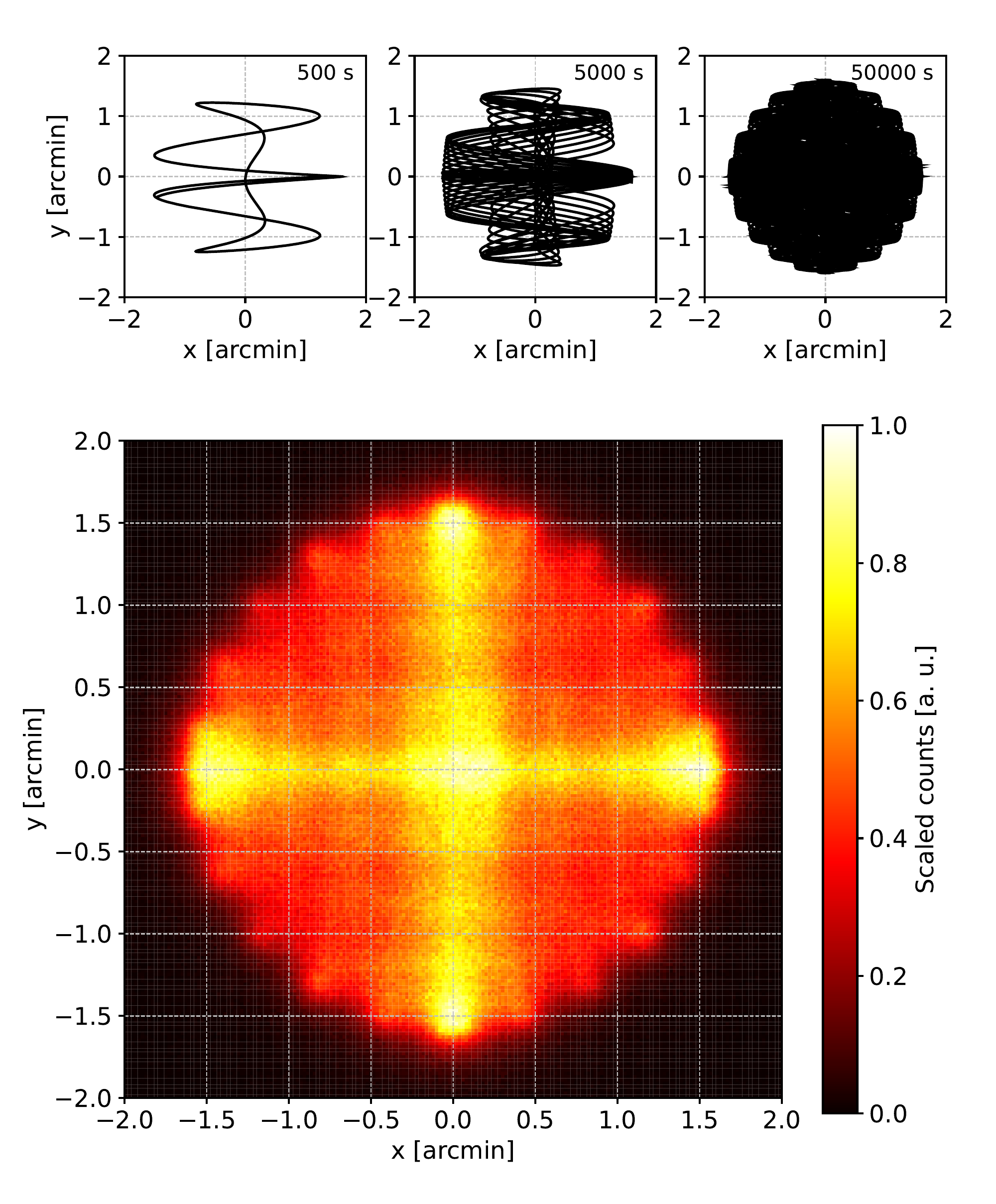}
    \caption{Representation of the default dithering pattern. The three small panels on the 
    top represent the dithering path around the target position for 500, 5000 and 50000~s, 
    while the histogram on the bottom represent the normalized counts after the convolution 
    with the PSF of the telescope.}
    \label{fig:dithering_pattern}
\end{figure}

As explained in~\cite{BALDINI2021102628}, the IXPE focal plane detectors feature 
systematic deviations from a flat azimuthal response to unpolarized radiation, characterized
by variations over small spatial scales, that need to be accounted for to reach the 
design polarization sensitivity. In order to average out this spurious modulation over the
detector surface and make the correction practically viable~\cite{Rankin_2022}, the
observatory is dithered around the pointing direction.
The dithering pattern has the form of a Lissajous figure with a circular envelope
\begin{align}
    \begin{cases}
    \delta x = a \cos(\omega_a t) \, \cos(\omega_x t) \\
    \delta y = a \sin(\omega_a t) \, \sin(\omega_y t)
    \end{cases}
\end{align}
with a default amplitude of $a = 1.6$~arcmin and the three periods corresponding to the
angular pulsations $\omega_a$, $\omega_x$ and $\omega_y$ being 907~s, 101~s and 449~s,
respectively (see Figure~\ref{fig:dithering_pattern}).

While the effect of the dithering is removed in the ground processing pipeline through the
knowledge of the aspect solution, and does not affect the source image in sky coordinates, 
the specifics of the observation strategy need to be captured by the simulation in order to
reproduce the morphology of the energy flux in detector coordinates (which is in turn
relevant for some instrumental effects) and for a correct calculation of the exposure. 
\ixpeobssim\ keeps track of the effect of the dithering and stores the pointing history,
sampled on a fixed-step, user-selectable, time grid in the (optional) \col{SC\_DATA} extension.

\section{Analysis Tools}
\label{sec:analysis}

\ixpeobssim\ comes with a set of facilities for binning event lists in several different
flavors. From an architectural standpoint, each binning algorithm comes with its own
interface classes for output (i.e., creating binned files from event lists) and input 
(i.e., reading, visualizing and manipulating binned FITS files).

\subsection{Basic polarization analysis}
\label{sec:pol_analysis}

The simplest approach that \ixpeobssim\ provides for polarization analysis is largely
borrowed from the model-independent approach described in~\cite{KISLAT201545}.
More specifically, for each event we define the three additive reconstructed quantities
\begin{align}\label{eq:polcubes}
    \begin{cases}
    \displaystyle \tilde{i_i} = \frac{w_i}{\aeff(E_i)}\\
    \displaystyle \tilde{q_i} = \frac{w_i q_i}{\aeff(E_i) \modf(E_i)}\\
    \displaystyle \tilde{u_i} = \frac{w_i u_i}{\aeff(E_i) \modf(E_i)},
    \end{cases}
\end{align}
where $w_i$ represents the (optional) event weights introduced in 
section~\ref{sec:introduction}. 
The on-axis effective area term in equation~\eqref{eq:polcubes} acts as an acceptance
correction guaranteeing that the relevant quantities are summed over a proxy of the input
source spectrum, as opposed to the measured count spectrum; note that $q_i$ and $u_i$ 
need to be divided by the proper modulation factor to transform the detector modulation
into the  actual polarization of the source.

We emphasize that the effective area and modulation factor in equation~\Eqref{eq:polcubes}
are calculated by default at the measured energy, i.e., the effect of the energy dispersion
is neglected. Experience shows that the effect is generally small, but \ixpeobssim\ allows
to verify it on a case-by-case basis, by using the ground truth for reference.

\begin{figure}[htb!]
    \centering
    \includegraphics[width=\linewidth]{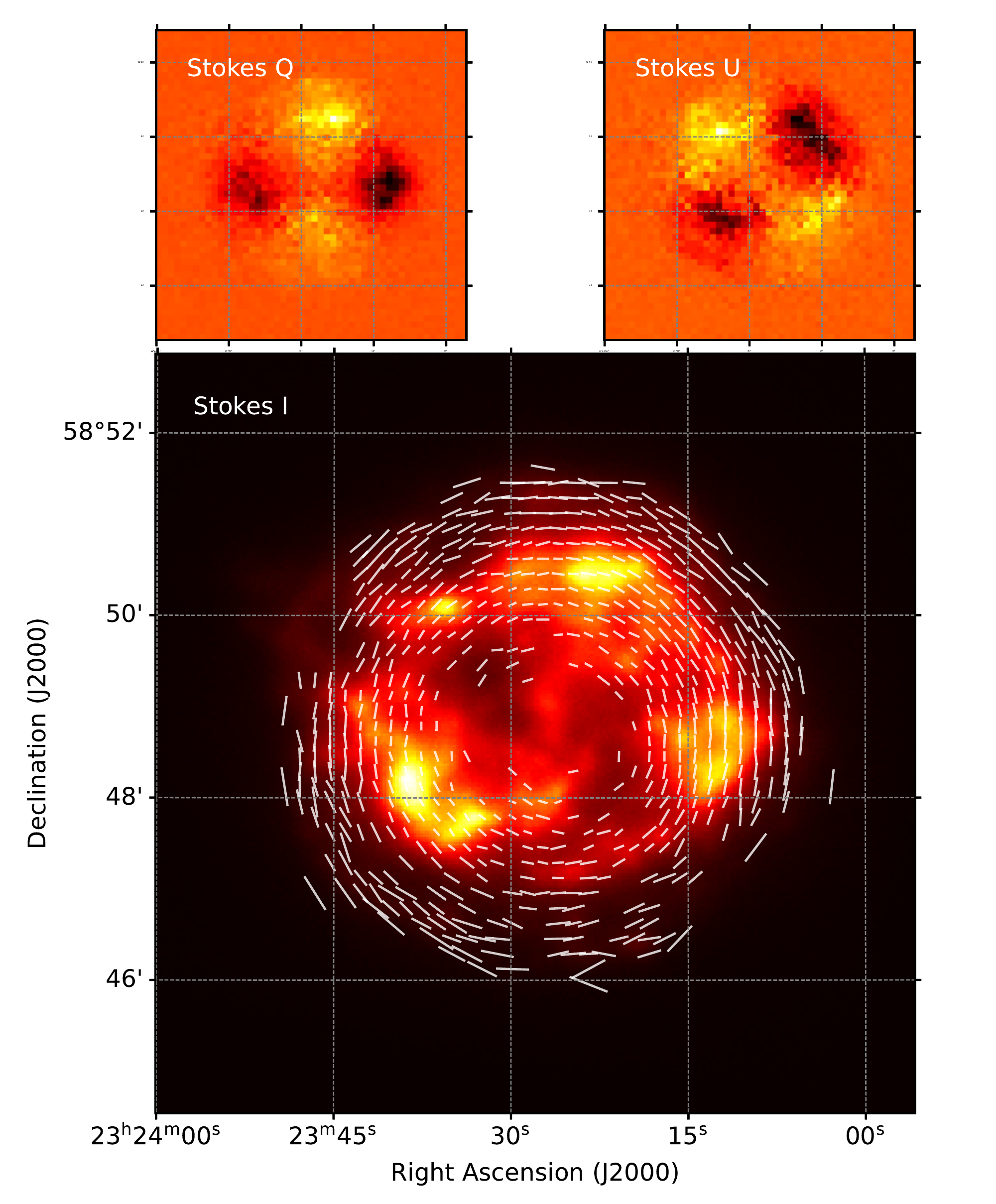}
    \caption{Polarization maps in the 2--8~keV energy band for a simulated, 2~Ms Cassiopea A
    observation, with a simple composite model including a thermal, un-polarized component,
    and a non-thermal, polarized component based on a simple geometric tangential pattern.
    (As it turned out after the actual observation of the source, the polarization degree
    of the model is largely unrealistic, but the figure is provided here purely for illustration
    purposes).
    The two panels on the top show the maps of the $Q$ and $U$ Stokes parameters in sky
    coordinates, while the main panel is a count map with the polarization direction for the
    pixels with a significance larger than $3\sigma$ overlaid. (The length of the arrows
    is proportional to the measured polarization degree.)}
    \label{fig:casa_polarization_map}
\end{figure}

The measured Stokes parameters over a generic subset $\mathcal{S}$ of the events (be that a
specific energy range, or a spatial bin in sky coordinates), is obtained by simply summing
the event-by-event quantities over $\mathcal{S}$. The polarization degree and angle can be
recovered with the usual formul\ae, and the formalism to propagate the statistical
uncertainties is thoroughly described in~\cite{KISLAT201545}. \ixpeobssim\ provides
facilities to calculate the broadband polarization properties over an arbitrary energy
binning, integrated over a given sub-region of the field of view, or binned in the sky, as
illustrated in~Figure~\ref{fig:casa_polarization_map}.
We note that, when coupled to a suitable minimizer, the \ixpeobssim\ convolution
capabilities could be effectively exploited to fit arbitrary spectro-polarimetric parametric
models for extended sources to a given observation---and this possibility is being 
actively investigated in the form a \package{ThreeML}~\cite{2015arXiv150708343V} 
plugin currently under development.

Among the additional analysis tools that are impossible to cover in the limited scope 
of this paper, we mention in passing \app{xpstokesalign}, that allows to align the Stokes
parameters, on an event-by-event basis, to a given polarization model, facilitating the
search for large-scale polarization signatures (e.g., radial or tangential) in extended
sources.

\subsection{Spectro-polarimetric fitting} 

\app{xpbin} provides dedicated algorithms to create spectra of Stokes parameters, binned in
pulse invariant channels, that can be readily used in conjunction with the standard fitting
tools used by the X-ray community, e.g., \package{XSPEC}~\cite{1996ASPC..101...17A},
\package{ThreeML}~\cite{2015arXiv150708343V} and \package{Sherpa}~\cite{2001SPIE.4477...76F},
to perform spectro-polarimetric fits~\cite{Strohmayer_2017}.
More specifically, \app{xpbin} can write standard PHA type-I files (with specific header
keywords for polarization analysis) containing the relevant binned quantities---that in the 
unweighted flavor read
\begin{align}
    \begin{cases}
    \displaystyle I_k = \frac{N_k}{T} \quad &
    \sigma_{I_k} = \displaystyle \frac{\sqrt{N_k}}{T}\\
    \displaystyle Q_k = \frac{1}{T} \sum_{\text{PI}=k} q_i \quad & 
    \sigma_{Q_k} = \displaystyle \frac{1}{T} \sqrt{\sum_{\text{PI}=k} q_i^2}\\
    \displaystyle U_k = \frac{1}{T} \sum_{\text{PI}=k} u_i \quad &
    \sigma_{U_k} = \displaystyle \frac{1}{T} \sqrt{\sum_{\text{PI}=k} u_i^2}.
    \end{cases}
\end{align}
It is important to notice that the binned spectra follow a pure counting statistics only
for the $I$ Stokes parameter in the un-weighted case, which has non trivial implications
in specific areas, such as the choice of a proper fitting statistics and/or of the optimal 
grouping algorithm for rebinning data where necessary.

We also emphasize that $I_k$, $Q_k$ and $U_k$ are expressed in detector space, and 
the detector response is taken into account by setting the proper response matrix and
ancillary response files---the effective area for the $I$ and the modulation response 
function for $Q$ and $U$.
The \ixpeobssim\ pseudo-CALDB provides response functions in both weighted and un-weighted 
fashion; a few simple, multiplicative polarimetric models are provided by HEASARC through
the page hosting XSPEC additional models%
\footnote{\url{https://heasarc.gsfc.nasa.gov/docs/xanadu/xspec/newmodels.html}}%
, and shipped with \ixpeobssim\ for convenience.

\subsection{Interface to \package{XSPEC}}

\begin{figure}[htb!]
    \centering
    \includegraphics[width=\linewidth]{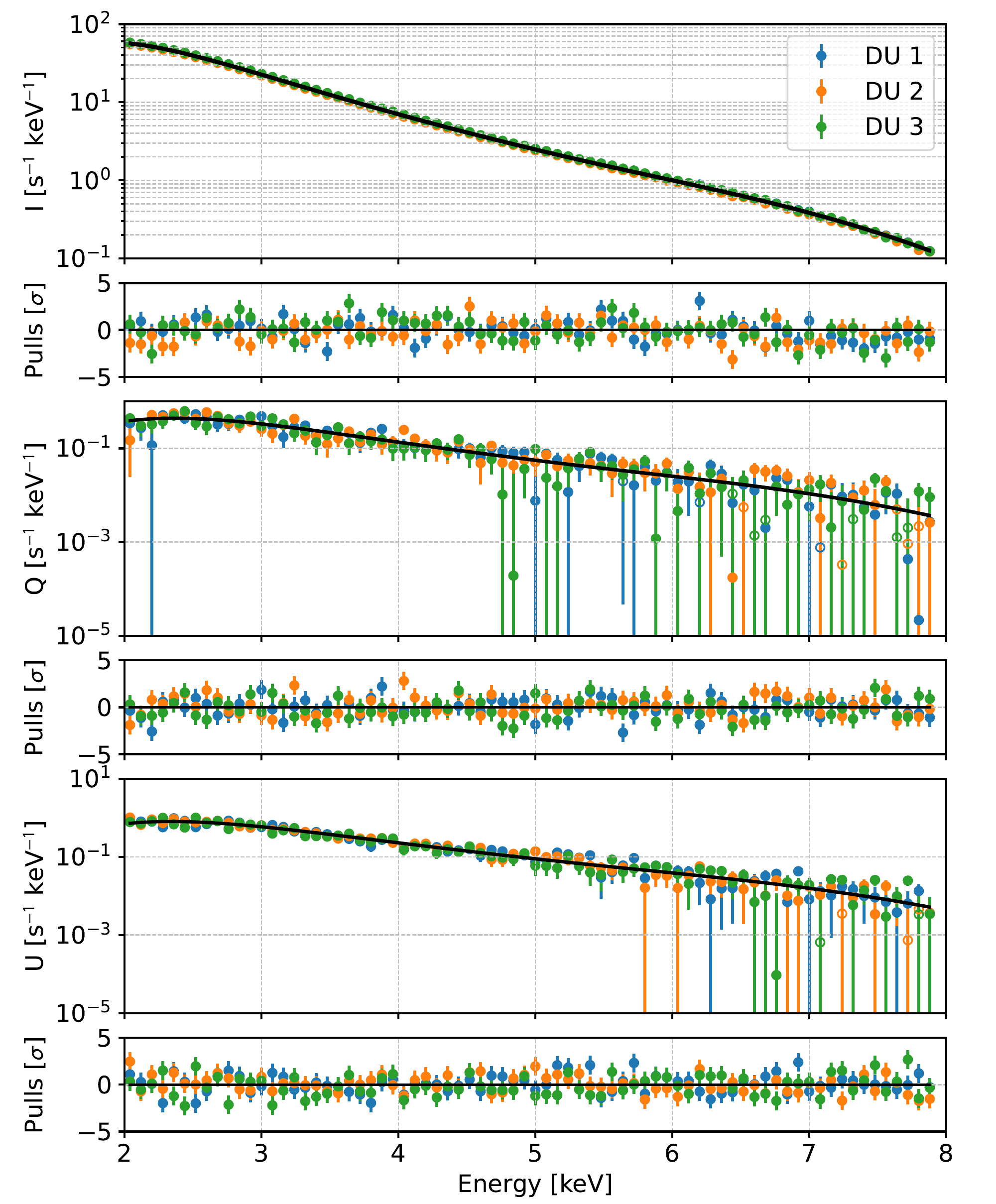}
    \caption{Spectro-polarimetric fit in \package{XSPEC} to simulated data of a 
    point source with a power-law spectrum and constant polarization degree and angle.
    This is a simultaneous, combined fit to 9 independent data sets (I, Q and U for each of 
    the three detector units) using a \texttt{pollin * powerlaw} model. The empty markers 
    in the $Q$ and $U$ spectra represent negative values, that could not otherwise be 
    rendered in logarithmic scale.}
    \label{fig:xspec_fit}
\end{figure}

Although the \app{xpbin} output can be used directly in \package{XSPEC} with the proper
response files, \ixpeobssim\ provides a lightweight Python wrapper, dubbed \app{xpxspec}
that facilitates combined spectral and spectro-polarimetric fits using the full data set
from the three IXPE detector units. Figure~\ref{fig:xspec_fit} shows an example of such a 
combined fit for a simulated point source with a power-law spectrum and a constant 
polarization degree and angle, displayed using the \ixpeobssim\ visualization facilities.

\subsection{Analysis pipelines}

As mentioned in the previous sections, one of the \ixpeobssim\ design goals since the very
beginning was to allow the user to develop simulation and analysis pipelines with minimal
effort. To this end, every single \ixpeobssim\ application is wrapped into a dedicated,
top-level module so that it can be effectively called from within a generic Python script
with the exact same keyword arguments that one would pass to the command line version of the
same script. These wrappers typically return the list of all the files that the function
call has created, which makes it very easy to chain application calls one after the other. 
The user is referred to the documentation for more information on this functionality, that
we deem as one of the most powerful of the entire framework, and a basic example is 
provided in~\ref{app:code}.

\section{Conclusions}

To support preparation for the IXPE mission, \ixpeobssim\ was developed to support 
advanced simulation and analysis facilities. With the IXPE data now being regularly delivered
to the public, we decided to change our development model and release the codebase under
an OSI-approved license, with the twofold purpose of benefiting the community engaged in
the data analysis and encourage reuse for future X-ray missions.

We emphasize that \ixpeobssim\ is under active development. In addition to the lack of
support for event weights discussed in~section~\ref{sec:weights} we are aware of a number of 
additional limitations, and we do have a clear path forward in most of the cases.
The relevant areas where we have room for improvement include the instrument description (e.g.,
the current simplistic, azimuthally-symmetric model for the PSF), the analysis algorithms in general,
as well as the definition of the public interfaces and their alignment with the publicly available
external tools.

We anticipate \ixpeobssim\ will be a useful tool for supporting the Guest Observing 
program. In addition, the ability to easily run nearly-identical analysis pipelines 
(including up-to-date calibration products) to both simulated and real IXPE observations
will enhance the ability of the scientific community to interpret this new and
complex frontier of X-ray observations.

\subsection*{Acknowledgements}

The Italian contribution to the IXPE mission is supported by the Italian Space Agency (ASI)
through the contract ASI-OHBI-2017-12-I.0, the agreements ASI-INAF-2017-12-H0 and
ASI-INFN-2017.13-H0, and its Space Science Data Center (SSDC), and by the Istituto Nazionale
di Astrofisica (INAF) and the Istituto Nazionale di Fisica Nucleare (INFN) in Italy.

This work was supported by the EU Horizon 2020 Research and Innovation Program under the Marie
Sklodowska-Curie Grant Agreement 734303.

We gratefully acknowledge members of the IXPE Collaboration for stimulating 
discussions and contributions to this work.

\bibliographystyle{unsrt}
\bibliography{ixpeobssim}

\onecolumn
\appendix

\section{Code Snippets}
\label{app:code}

\begin{lstlisting}[language=Python, 
caption={Minimal example of a configuration file for a field containing a single, stationary 
point source with a power-law photon spectrum and a polarization degree linearly increasing
with energy (the corresponding position angle is constant, and aligned with the celestial North).}
]
import numpy

from ixpeobssim.srcmodel.bkg import xTemplateInstrumentalBkg
from ixpeobssim.srcmodel.polarization import constant
from ixpeobssim.srcmodel.roi import xPointSource, xROIModel
from ixpeobssim.srcmodel.spectrum import power_law

# Sky position and spectral parameters of the source.
SRC_RA, SRC_DEC = 20., 30.
PL_NORM = 6.e-3
PL_INDEX = 2.
# The pointing direction is the same as the source coordinates.
PNT_RA, PNT_DEC = SRC_RA, SRC_DEC

# Definition of the photon spectrum.
spec = power_law(PL_NORM, PL_INDEX)

def pol_deg(E, t=None, ra=None, dec=None):
    """Definition of the polarization degree as a function of the energy.

    The polarization degree is 5% at 1 keV, increasing linearly with energy.
    """
    return 0.05 * E

# Definition of the polarization angle---0. is aligned with the North.
pol_ang = constant(0.)

# Definition of the sources and the region of interest.
src = xPointSource('Point source', SRC_RA, SRC_DEC, spec, pol_deg, pol_ang)
bkg = xTemplateInstrumentalBkg()
ROI_MODEL = xROIModel(PNT_RA, PNT_DEC, src, bkg)
\end{lstlisting}

\clearpage

\begin{lstlisting}[language=Python, 
caption={Sample simulation and analysis pipeline for the example.
This will run a simulation for the specified configuration file, select
data for the source and the background regions, and perform a model-independent
polarization analysis in a series of energy bins.}
]
import os

import numpy

from ixpeobssim import IXPEOBSSIM_CONFIG
import ixpeobssim.core.pipeline as pipeline
from ixpeobssim.binning.polarization import xBinnedPolarizationCube

# Basic simulation parameters.
CFG_FILE_PATH = os.path.join(IXPEOBSSIM_CONFIG, 'toy_softwarex.py')
DURATION = 2000000.
# Global flag---toggle this not to overwrite existing files.
OVERWRITE = True
# Region selection: the source is a circular patch, while the background is a larger annulus centered
# in the same position (by default the reference position in the WCS of the original event file).
# All radii are in arcmin.
SRC_RAD = 0.75
BKG_INNER_RAD = 1.5
BKG_OUTER_RAD = 3.
# Energy binning for the polarization cubes.
ENERGY_BINNING = numpy.array([2., 4., 6., 8.])

# Run the simulation.
file_list = pipeline.xpobssim(configfile=CFG_FILE_PATH, duration=DURATION, overwrite=OVERWRITE)

# Select the source and the background regions. Note this will keep track of the area for each
# selection by setting the BACKSCAL header keyword in the output file.
src_file_list = pipeline.xpselect(*file_list, rad=SRC_RAD, suffix='src', overwrite=OVERWRITE)
bkg_file_list = pipeline.xpselect(*file_list, innerrad=BKG_INNER_RAD, rad=BKG_OUTER_RAD,  
    suffix='bkg', overwrite=OVERWRITE)

# Create the polarization cubes.
kwargs = dict(algorithm='PCUBE', ebinalg='LIST', ebinning=ENERGY_BINNING, overwrite=OVERWRITE)
src_pcube_file_list = pipeline.xpbin(*src_file_list, **kwargs)
bkg_pcube_file_list = pipeline.xpbin(*bkg_file_list, **kwargs)

# Read back the polarization cubes and perform the background subtraction.
src_pcube = xBinnedPolarizationCube.from_file_list(src_pcube_file_list)
bkg_pcube = xBinnedPolarizationCube.from_file_list(bkg_pcube_file_list)
bkg_pcube *= src_pcube.backscal() / bkg_pcube.backscal()
src_pcube -= bkg_pcube

# You are good to go!
print('Polarization degree : ', src_pcube.PD)
print('Polarization degree error : ', src_pcube.PD_ERR)
print('Polarization angle : ', src_pcube.PA, 'deg')
print('Polarization angle error : ', src_pcube.PA_ERR, 'deg')

\end{lstlisting}

\end{document}